\documentstyle[preprint,aps]{revtex}
\tightenlines

\begin{document}
\begin{center}
{\Large \bf Superposition of chaotic process with convergence to L\'evy's stable law}\\ 
\vskip.25in
{ Ken Umeno}

 {\it BSI, The Institute of Physical 
  and Chemical Research (RIKEN)\\ 
2-1 Hirosawa, Wako, Saitama 351-01, Japan} 
\vskip.25in
 (Received: October 23, 1997)
\end{center}   
\begin{abstract}
We construct a family of 
chaotic dynamical systems with explicit broad distributions,
which always violate  the central limit theorem.
In particular,
we show that 
the superposition of many statistically independent, identically distributed random 
variables obeying such chaotic process converge in density to 
L\'evy's stable laws in a full range of the index parameters.
The theory related to the connection between 
deterministic chaos and non-Gaussian distributions  
gives us a systematic view of the purely mechanical generation of 
L\'evy's stable laws.
\end{abstract}
\clearpage
\setcounter{equation}{0}
The central limit theorem (CLT) breaks down when any stationary stochastic process has infinite variance.
Thus, it is an important question whether this mathematically pathological situation is physically 
relevant or not.
One noteworthy point arising from recent studies is that  
such examples of L\'evy's stable laws, which are the most famous class of distributions
  violating the CLT, can be seen in 
 many different fields, such as astronomy,
physics, biology, economics and communication engineering, 
under broad conditions\cite{holtzmark,mandelbrot,Mantegna,bouchaud}. 
Thus, it is our primary interest to determine why such L\'evy's stable laws 
are widely observed, and 
  furthermore, to elucidate the mechanism of generating 
L\'evy's stable laws. 
In the 1980's, there were several studies which clarified the relation between 
 intermittent periodic 
mapping   and anomalous diffusion with L\'evy's law-like broad distributions\cite{man,geisel}. 
Random-walk models\cite{Hughes}
and combinations of 
several random number generators\cite{devroye,Mantegena2} 
are also utilized to generate L\'evy's stable laws. 
However, such analyses include approximations or the non-deterministic 
nature in 
the models themselves, or else their generation methods are 
only applicable  to a special class 
of L\'evy's stable laws.
The purpose of the present paper is to present a 
systematic method for exact and purely 
mechanical generation of stable laws with arbitrary indices only 
 using concrete chaotic dynamical  systems. 
Let us consider an one-dimensional dynamical system
\begin{equation} 
\label{eq:cauchy}
X_{n+1}=\frac{1}{2}(X_{n}-\frac{1}{X_{n}})\equiv f(X_{n})
\end{equation}
on the infinite support \((-\infty,+\infty)\).
 Note that 
this mapping \(f(X)\) 
can be seen as the doubling formula of \(-\cot (\theta)\) as
\(-\cot(2\theta)=f[-\cot (\theta)]\). Thus, 
the system has the exact solution  
\(X_{n}=-\cot(\frac{\pi}{2}2^{n}\theta_{0})\).    
 Using a diffeomorphism
\(x\equiv\phi^{-1}(\theta)=
-\frac{1}{\tan(
\frac{\pi}{2}\theta)}\)
 of \(\theta\in [0,2]\) into \(]-\infty,+\infty[\),
 we derive the piecewise-linear map  
\(g^{(2)}(\theta)=\phi \circ f
\circ \phi^{-1} (\theta)\) as 
\begin{equation}
\label{eq:piece_2}
\begin{array}{l}
g^{(2)}(\theta)=2\theta,\quad \theta\in [0,1)\\
g^{(2)}(\theta)=2\theta-2,\quad \theta\in [1,2].
\end{array}
\end{equation}
Because the map (\ref{eq:piece_2}) has the mixing property (thus, is clearly  
 ergodic) and preserves the Lebesgue measure \(\frac{1}{2}d\theta\)
of \([0,2]\), the map \(f\) preserves the measure
\begin{equation}
\label{eq:cauchy_dist}
 \mu(dx)=\rho(x)dx=
\frac{1}{2}\frac{d\phi(x)}{dx}dx=\frac{dx}{\pi(1+x^{2})}.
\end{equation}
This is an explanation of the 
mechanical  origin of the Cauchy distribution (\ref{eq:cauchy_dist}). 
Note that  the Cauchy distribution is 
a simple case of L\'evy stable distributions with the characteristic 
\(\alpha=1\). The measure (\ref{eq:cauchy_dist}) is absolutely continuous 
 with respect to the Lebesgue measure, which implies that the Kolmogorov-Sinai
entropy \(h(\mu)\) is equivalent to the Lyapunov exponent of ln2 from the 
Pesin identity; and the measure is a physical one in the sense that, for 
almost all initial conditions \(x_{0}\), the time averages 
\(\lim_{n\rightarrow \infty}(1/n)\sum_{i=0}^{n-1}\delta(x-x_{i})\) reproduce
the invariant measure \(\mu(dx)\)\cite{eckmann}. 
Our  next step is to         
generalize exactly solvable chaos (\ref{eq:cauchy}) to capture 
the full domain of Levy's stable laws. 
Now, let us consider the  mapping 
\begin{equation}
\label{eq:general_2}
X_{n+1}=|\frac{1}{2}(|X_{n}|^{\alpha}-1/|X_{n}|^{\alpha})|^{\frac{1}{\alpha}}
 \cdot \mbox{sgn}[(X_{n}-1/X_{n})]
\equiv f_{\alpha}(X_{n}),
\end{equation}
 where \(\mbox{sgn}(x)=1\) for \(x>0\) and \(\mbox{sgn}(x)=-1\) for 
\(x<0\).
We prove here that this chaotic dynamics (\ref{eq:general_2}) 
also has the mixing property similar to   mapping (\ref{eq:cauchy}),
as well as the
 exact invariant 
density function given by   
\begin{equation}
\label{eq:density_alpha}
   \rho_{\alpha}(x)=\frac{\alpha}{\pi}\frac{|x|^{\alpha-1}}{(1+|x|^{2\alpha})}
\simeq 
\frac{\alpha}{\pi}|x|^{-(\alpha+1)}\quad\mbox{for}
\quad |x|\rightarrow \infty.
\end{equation}
Note that the chaotic system (\ref{eq:cauchy}) is a special case 
of (\ref{eq:general_2}) with \(\alpha=1\). 
We remark here  that this system 
can also be seen as a doubling formula  
\(s(2\theta)=f_{\alpha}[s(\theta)]\), 
where  \(s(\theta)= -\frac{\mbox{sign}[\tan(\frac{\pi}{2}\theta)]}{|\tan(
\frac{\pi}{2}\theta)|^{1/\alpha}}\).

Using the relations
\begin{equation}
\begin{array}{l}
s(2\theta)=f_{\alpha}[s(\theta)]\quad \mbox{for}\quad \theta\in[0,1)\\
s(2\theta-2)=f_{\alpha}[s(\theta)]\quad \mbox{for}\quad \theta\in[1,2]\\
\end{array}
\end{equation}  
 and  defining the 
diffeomorphism
\begin{equation} 
x=\phi^{-1}_{\alpha}(\theta)=
-\frac{\mbox{sgn}[\tan(\frac{\pi}{2}\theta)]}{|\tan(
\frac{\pi}{2}\theta)|^{1/\alpha}} 
\end{equation} 
 of \(\theta\in [0,2]\) into \(]-\infty,+\infty[\), 
 we obtain the piecewise-linear map  
\(g^{(2)}(\theta)=\phi_{\alpha} \circ f_{\alpha} 
\circ \phi^{-1}_{\alpha} (\theta)\) as
\begin{equation}
\label{eq:piece_2_2}
g^{(2)}(\theta)=
\left\{
\begin{array}{l}
2\theta,\quad \theta\in [0,1)\\
2\theta-2,\quad \theta\in [1,2]
\end{array}
\right.
\end{equation}
with the invariant measure \(\frac{1}{2}d\theta\) of \([0,2]\). Thus,
map (\ref{eq:general_2}) preserves 
\begin{equation}
  \mu(dx)=\rho_{\alpha}(x)dx=\frac{1}{2}\frac{d\phi_{\alpha}}{dx}dx=
  \frac{\alpha}{\pi}\frac{|x|^{\alpha-1}}{(1+|x|^{2\alpha})}.
\end{equation}

Therefore, 
the class of dynamical systems (\ref{eq:general_2}) with the  parameter
\(\alpha\) also has a mixing property (thus, is ergodic) with the Lyapunov 
exponent ln2.

More generally, from the family of Chebyshev maps 
\(Y_{n+1}=f(Y_{n})\) defined by the addition formulas of the form 
\(\sin^{2}(p\theta)=f[\sin^{2}(\theta)]\), where \(p=2,3,\cdots,\) with 
the unique density \(\sigma(y)=\frac{1}{\pi\sqrt{y(1-y)}}\) of the logistic 
map \(Y_{n+1}=4Y_{n}(1-Y_{n})\) (which corresponds to the case \(p=2\))\cite{uf,adler}, 
we may construct
infinitely many  chaotic dynamical systems 
\(X_{n+1}=f_{\alpha}(X_{n})\) with the unique density function 
\(\rho_{\alpha}(x)\) given by Eq.(\ref{eq:density_alpha})\cite{ku0}. 
For example, an explicit mapping  with the Lyapunov exponent ln3
is given by 
\begin{equation}
X_{n+1}=f_{\alpha}(X_{n})=\mid\frac{|X_{n}|^{\alpha}(|X_{n}|^{2\alpha}-3)}
{(3|X_{n}|^{2\alpha}-1)}\mid^{\frac{1}{\alpha}}\cdot \mbox{sign}
[\frac{X_{n}(|X_{n}|^{2\alpha}-3)}{(3|X_{n}|^{2\alpha}-1)}], 
\end{equation}
 which has the density (\ref{eq:density_alpha}) from the triplication formula 
of \(s(\theta)\).
 
In this case, the  topological conjugacy relation 
\(g^{(3)}(\theta)=\phi_{\alpha}\circ f_{\alpha}\circ 
\phi_{\alpha}^{-1}(\theta)\)  yields 
the piecewise-linear map: 
\begin{equation}
g^{(3)}(\theta)=
\left\{
\begin{array}{l}
3\theta,\quad \theta\in [0,\frac{2}{3})\\
3\theta-2,\quad \theta\in [\frac{2}{3},\frac{4}{3})\\
3\theta-4,\quad \theta\in [\frac{4}{3},2].
\end{array}
\right.
\end{equation}
In general, the same kind of topological conjugacy relation 
\(g^{(p)}(\theta)=\phi_{\alpha}\circ f_{\alpha}\circ 
\phi_{\alpha}^{-1}(\theta)\) holds for 
a \(p\)-to-one piece-wise linear mapping \(g^{(p)}(\theta)\). 
Let us consider 
slightly modified dynamical 
systems \(X_{n+1}=f_{\alpha,\delta}(X_{n})\equiv 
\frac{1}{\delta}f_{\alpha}(\delta X_{n})\) with a change of variable 
\(h(x)\equiv \delta x\) for a constant
\(\delta>0\).  Thus, this modified dynamics,  
\begin{equation}
 f_{\alpha,\delta}(X)=
|\frac{1}{2}(|X|^{\alpha}-1/|\delta^{2}X|^{\alpha})|^{\frac{1}{\alpha}}
 \cdot \mbox{sgn}[X-\frac{1}{\delta^{2} X}],
\end{equation}
has an invariant measure 
\(\rho_{\alpha,\delta}(x)dx=\delta\rho_{\alpha}(\delta x)dx=
\frac{\alpha\delta^{\alpha}|x|^{\alpha-1}dx}
{\pi(1+\delta^{2\alpha}|x|^{2\alpha})}\) 
with a slightly modified power-law tail as 
\begin{equation}
\rho_{\alpha,\delta}(x)\simeq \frac{\alpha}{\pi\delta^{\alpha}}
\frac{1}{|x|^{\alpha+1}}\quad \mbox{for}\quad x\rightarrow \pm\infty.
\end{equation}
We will show that this power-law tail of density is sufficient for 
generating arbitrary symmetric stable laws.   
The canonical representation of stable laws  
obtained by L\'evy and Khintchine\cite{Khintchine,levy} is:
\begin{equation}
 P(x;\alpha,\beta)=\frac{1}{2\pi}\int_{-\infty}^{\infty}\exp(izx)
\psi(z)dz,
\end{equation}
 where the characteristic function \(\psi(z)\) is given by 
\begin{equation}
  \psi(z)=\exp\{-i\gamma z-\eta |z|^{\alpha}[1+i\beta\mbox{sgn}(z)
\omega(z,\alpha)]\},
\end{equation}
\(\alpha,\beta,\gamma\) and \(\eta\) being  real constants satisfying 
 \( 0<\alpha\leq 2,-1\leq \beta \leq 1,\gamma\geq0\)
and 
\begin{equation}
\begin{array}{l}
\omega(z,\alpha)=\tan(\pi\alpha/2)\quad \mbox{for}\quad \alpha\ne1,\\
\omega(z,\alpha)=(2/\pi)\log|z|\quad \mbox{for}\quad \alpha=1.
\end{array}
\end{equation}
According to the generalized central limit theorem (GCLT),  
it is known\cite{gnedenko} that 
if the density function of a stochastic  process has a  
long tail, 
\begin{equation}
\rho(x)\simeq c_{-}|x|^{-(1+\alpha)}\quad {\mbox for}\quad 
  x\rightarrow -\infty,\quad 
\rho(x)\simeq c_{+}|x|^{-(1+\alpha)}\quad {\mbox for}\quad 
  x\rightarrow +\infty,
\end{equation}
then the superposition \(S_{N}=(\sum_{i=1}^{N}X(i)-A_{N})/B_{N}\) 
of independent, identically distributed random variables \(X(1),\cdots, X(N)\) 
with the density \(\rho(x)\)  
converges in density to a L\'evy's stable law \(P(x;\alpha,\beta)\) with 
\begin{equation}
\begin{array}{l}
 \beta=(c_{+}-c_{-})/(c_{+}+c_{-}),\\
  A_{N}=0,B_{N}=N^{1/\alpha},\eta=\frac{\pi(c_{+}+c_{-})}{2\alpha\sin(\pi\alpha/2)\Gamma(\alpha)},\quad 
\mbox{for}\quad 0<\alpha<1,\\
 A_{N}=N<x>,B_{N}=N^{1/\alpha},\eta=\frac{\pi(c_{+}+c_{-})}
{2\alpha^{2}\sin(\pi\alpha/2)\Gamma(\alpha-1)},
\quad \mbox{for}\quad 1<\alpha<2.
\end{array}
\end{equation}

In the case of the symmetric long-tail (\ref{eq:density_alpha}),
 \(c_{+}=c_{-}=\frac{\alpha}{\pi\delta^{\alpha}}\),\(\beta=0\) and 
 \(\eta\)  is 
determined by 
\begin{equation}
\begin{array}{l}
\eta=\frac{1}{\delta^{\alpha}\sin(\frac{\pi\alpha}{2})\Gamma(\alpha)}\quad
\mbox{for}\quad 0<\alpha<1,\\
\eta=\frac{1}{\alpha\delta^{\alpha}\sin(\frac{\pi\alpha}{2})\Gamma(\alpha-1)}
\quad \mbox{for}\quad 1<\alpha<2.
\end{array}
\end{equation}

Thus, according to the GCLT, the superposition of statistically independent, identically distributed 
random variables generated by  \(N\) unique chaotic systems 
(\ref{eq:general_2}) 
is guaranteed to converge in distribution to an arbitrary 
 symmetric L\'evy's stable law \(P(x;\alpha,\beta=0)\).
Figure 1 shows that   the 
convergence in distribution to a L\'evy's stable distribution with 
parameters \(\alpha=1.5\) and \(\beta=0\) is clearly seen for \(N=10000\), 
as predicted by the GCLT. 
What about more general stable distributions including 
asymmetric stable laws? 
In the next part, we will also provide  chaotic 
dynamical systems with explicit broad distributions, whose superpositions converge in distribution to arbitrary 
asymmetric L\'evy's stable laws. 
Let us consider a family of dynamical systems 
\(X_{n+1}=f_{\alpha,\delta_{1},\delta_{2}}(X_{n})\), where 
\begin{equation}
\label{eq:general_f}
f_{\alpha,\delta_{1},\delta_{2}}(X)=
\left\{ \begin{array}{l}
  \frac{1}{\delta_{1}^{2}|X|}(\frac{|\delta_{1}X|^{2\alpha}
   -1}{2})^{\frac{1}{\alpha}}
   \quad \mbox{for}\quad X>\frac{1}{\delta_{1}}\\
-\frac{1}{\delta_{1}\delta_{2}|X|}
 (\frac{1-|\delta_{1}X|^{2\alpha}}{2})^{\frac{1}{\alpha}}  
  \quad  
  \mbox{for}\quad 0<X<
   \frac{1}{\delta_{1}}\\
\frac{1}{\delta_{1}\delta_{2}|X|}
 (\frac{1-|\delta_{2}X|^{2\alpha}}{2})^{\frac{1}{\alpha}}
\quad
  \mbox{for}\quad  
  -\frac{1}{\delta_{2}}<X<0\\
-\frac{1}{\delta_{2}^{2}|X|}(\frac{|\delta_{2}X|^{2\alpha}
   -1}{2})^{\frac{1}{\alpha}}
\quad \mbox{for}\quad X<-\frac{1}{\delta_{2}}.
\end{array}
\right.
\end{equation}
We can show 
that the dynamical system \(X_{i+1}=f_{\alpha,\delta_{1},\delta_{2}}(X_{i})\)
 has an asymmetric 
invariant measure, \(\mu(dx)=\rho(x;\alpha,\delta_{1},\delta_{2})dx\), 
where \(\rho_{\alpha,\delta_{1},\delta_{2}}(x)\) is given by 
\begin{equation}
\label{eq:density_asymmetric}
\rho_{\alpha,\delta_{1},\delta_{2}}(x)=
\frac{\alpha\delta_{1}^{\alpha}x^{\alpha-1}}
{\pi(1+\delta_{1}^{2\alpha}x^{2\alpha})}\quad \mbox{for}\quad x>0,\quad
\rho_{\alpha,\delta_{1},\delta_{2}}(x)=
\frac{\alpha\delta_{2}^{\alpha}|x|^{\alpha-1}}
{\pi(1+\delta_{2}^{2\alpha}|x|^{2\alpha})}\quad \mbox{for}\quad x<0.
\end{equation}  
Because the power-law tail is asymmetric as
\begin{equation}
\rho_{\alpha,\delta_{1},\delta_{2}}(x)
\simeq \frac{\alpha}{\pi
\delta_{1}^{\alpha}x^{\alpha+1}}\quad x\rightarrow +\infty,\quad
\rho_{\alpha,\delta_{1},\delta_{2}}(x)
\simeq \frac{\alpha}{\pi
\delta_{2}^{\alpha}|x|^{\alpha+1}}\quad x\rightarrow -\infty
\end{equation}
for \(\delta_{1}\ne \delta_{2}\),
the GCLT guarantees that
the limiting distribution would be a L\'evy's canonical form 
 \(P(x;\alpha,\beta)\) with the skewness parameter
\(\beta=\frac{\delta_{2}^{\alpha}-\delta_{1}^{\alpha}}{\delta_{1}^{\alpha}
 +\delta_{2}^{\alpha}}\ne 0\). Thus, we can generate arbitrary 
L\'evy's stable laws \(P(x;\alpha,\beta)\) for \(0<\alpha\leq 2\) and 
\(-1\leq \beta \leq 1\)\cite{beta} using the chaotic mappings 
\(f_{\alpha,\delta_{1},\delta_{2}}(X)\) with proper parameters \(\alpha\),
\(\delta_{1}\) and \(\delta_{2}\). 
Convergence to an asymmetric L\'evy's stable 
distribution with indices \(\alpha=1.5\) and \(\beta=-\frac{9-4\sqrt{2}}{7}\)
is clearly seen for \(N=10000\) in Fig.2, as predicted by the GCLT.  
To show the exactness of the asymmetric density 
(\ref{eq:density_asymmetric}), we must check that 
the invariant measure \(\rho_{\alpha,\delta_{1},\delta_{2}}(x)dx\) 
satisfies  the probability preservation relation (Perron-Frobenius Equation)
\cite{lasota}: 
\begin{equation}
\label{eq:pf_2}
 \rho_{\alpha,\delta_{1},\delta_{2}}(y)=\sum_{x=f_{\alpha,\delta_{1},\delta_{2}}^{-1}(y)}
\rho_{\alpha,\delta_{1},\delta_{2}}(x).
|\frac{dx}{dy}|.  
\end{equation}
We note that \(f_{\alpha,\delta,\delta}(x)=f_{\alpha,\delta}(x)\) and 
\(\rho_{\alpha,\delta_{1},\delta_{2}}(x)=\rho_{\alpha,\delta_{1}}(x)\) for \(x>0\) and 
\(\rho_{\alpha,\delta_{1},\delta_{2}}(x)=\rho_{\alpha,\delta_{2}}(x)\) for \(x<0\), which also 
have
the Perron-Frobenius equations  
\begin{equation}
\label{eq:pf_1}
  \rho_{\alpha,\delta_{i}}(y)=\sum_{x=f_{\alpha,\delta_{i}}^{-1}(y)}
\rho_{\alpha,\delta_{i}}(x)|\frac{dx}{dy}|\quad \mbox{for}\quad
 i=1,2. 
\end{equation}
Here, we define two pre-images \(x_{a}=f^{-1}_{\alpha,\delta_{1},\delta_{2}}(y)<0\) and 
 \(x_{b}=f^{-1}_{\alpha,\delta_{1},\delta_{2}}(y)>0\) for  \(y>0\).
In the case \(f_{\alpha,\delta_{1}}(x)=y>0\), we also 
define two pre-images  
\(x'_{a}=f^{-1}_{\alpha,\delta_{1}}(y)<0\) and 
\(x'_{b}=f^{-1}_{\alpha,\delta_{1}}(y)(=x_{b})>0\) for 
\(y=f_{\alpha,\delta_{1}}(x)>0\).
It is easy to check 
that \(\delta_{2}x_{a}=\delta_{1}x'_{a}\).
From the Perron-Frobenius equations (\ref{eq:pf_2}) and (\ref{eq:pf_1}),
we have the relation
\begin{equation}
\label{eq:relation}
  \rho_{\alpha,\delta_{2}}(x_{a})\frac{1}{|\frac{df_{\alpha,\delta_{1},\delta_{2}}(x)}{dx}|_{x=x_{a}}}
  =\rho_{\alpha,\delta_{1}}(x'_{a})\frac{1}{|\frac{df_{\alpha,\delta_{1}}(x)}{dx}|_{x=x'_{a}}}.
\end{equation}  
However, we can check the  relation (\ref{eq:relation}) under the condition  \(\delta_{2}x_{a}=\delta_{1}x'_{a}\).
In the same manner,  we can also show that \(\rho_{\alpha,\delta_{1},\delta_{2}}(y)\) satisfies
the Perron-Frobenius equation (\ref{eq:pf_2}) for \(y<0\).

There is also an interesting dualistic structure in these types of 
chaotic dynamical systems.
Let us consider  dynamical systems 
\(X^{*}_{n+1}=f^{*}_{\alpha,\delta_{1},\delta_{2}}(X^{*}_{n})\) 
defined as  
\begin{equation}
\label{eq:ura}
f^{*}_{\alpha,\delta_{1},\delta_{2}}(X^{*})=
\left\{
\begin{array}{l}
-\frac{\delta_{1}}{\delta_{2}}
 (\frac{2|X^{*}|^{\alpha}}{|\delta_{1}X^{*}|^{2\alpha}-1})^{\frac{1}{\alpha}}
   \quad \mbox{for}\quad X^{*}>\frac{1}{\delta_{1}}\\
 (\frac{2|X^{*}|^{\alpha}}{1-|\delta_{1}X^{*}|^{2\alpha}})^{\frac{1}{\alpha}}
  \quad  
  \mbox{for}\quad 0<X^{*}<
   \frac{1}{\delta_{1}}\\
 -(\frac{2|X^{*}|^{\alpha}}{1-|\delta_{2}X^{*}|^{2\alpha}})^{\frac{1}{\alpha}}
  \mbox{for}\quad  
  -\frac{1}{\delta_{2}}<X^{*}<0\\
\frac{\delta_{2}}{\delta_{1}}
 (\frac{2|X^{*}|^{\alpha}}{|\delta_{2}X^{*}|^{2\alpha}-1})^{\frac{1}{\alpha}}
\quad \mbox{for}\quad X^{*}<-\frac{1}{\delta_{2}}.
\end{array}
\right.
\end{equation}
Because the normalized and symmetric  case, 
 \(f^{*}_{\alpha}(X^{*})\equiv f^{*}_{\alpha,\delta_{1}=1,
\delta_{2}=1}(X^{*})\),
can also be seen as the doubling formula of \(-\mbox{sgn}[\tan(\frac{\pi}{2}\theta)]
|\tan(\frac{\pi}{2}\theta)|^{1/\alpha}\), and 
we can derive the piecewise-linear map 
\(g^{(2)}(\theta)=\phi^{*}_{\alpha}\circ f^{*}_{\alpha}\circ \phi^{*-1}_{\alpha}(\theta)\) (\ref{eq:piece_2}) by the diffeomorphism
\(x^{*}=\phi_{\alpha}^{*-1}(\theta)=-\mbox{sgn}[\tan(\frac{\pi}{2}\theta)]
|\tan(\frac{\pi}{2}\theta)|^{1/\alpha}\),
 map \(f^{*}_{\alpha}(X^{*})\) also has the same invariant measure 
\(\rho_{\alpha}(x)\) given by  
Eq. (\ref{eq:density_alpha}). Thus, in the same way as that used in obtaining 
\(\rho_{\alpha,\delta_{1},\delta_{2}}(x)\) for 
\(f_{\alpha,\delta_{1},\delta_{2}}(X)\),  
we can show that  map (\ref{eq:ura}) has  
the  invariant measure \(\rho_{\alpha,1/\delta_{1},1/\delta_{2}}(x)dx\). 
Furthermore, 
the dynamical reciprocal relation 
\begin{equation}
f^{*}_{\alpha,1/\delta_{1},1/\delta_{2}}(X^{*})f_{\alpha,\delta_{1},\delta_{2}}(X)=1\quad \mbox{for}\quad
XX^{*}=1
\end{equation}
holds.
This dualistic structure of the dynamical systems \(f_{\alpha,\delta_{1},\delta_{2}}(X)\)
  and \(f^{*}_{\alpha,1/\delta_{1},1/\delta_{2}}(X^{*})\)  originates  from the relation 
 \(\phi^{-1}_{\alpha}(\theta)\cdot \phi^{*-1}_{\alpha}(\theta)=1\). 

 In summary,  
we have found that L\'evy's stable laws can be directly generated by the 
superposition of many independent, identically distributed dynamical 
variables obeying certain unique chaotic processes. 
Owing to the ubiquitous character of L\'evy's stable laws, 
our methods related to ergodic transformations with long-tail densities 
have a broad range of applications to 
many different  physical  problems where stable laws 
play an essential role.  
This work is  supported by RIKEN-SPRF and RIKEN-PSR grants.    

\clearpage 
\begin{large}
\begin{center}
FIGURES
\end{center}
\end{large}
FIG. 1.(color)\quad
Densities of the superposition \(S_{N}=
(\sum_{i=1}^{N}X(i)-A_{N})/B_{N}\) of dynamical variables \(X(i)\) generated 
by chaotic systems \(X_{j+1}(i)=
f^{(2)}_{\alpha=1.5}[X_{j}(i)]\), 
 with \(N\) 
different initial conditions \(X_{0}(i)|_{i=1,\cdots,N}\)
are plotted for 
\(N=1,10,100,1000\) and \(10000\). In this case, the limit density converges to the symmetric 
L\'evy's stable law with the indices \(\alpha=1.5\)  and \(\beta=0\).\\[1cm]
FIG. 2.(color)\quad
Densities of the superposition \(S_{N}=
(\sum_{i=1}^{N}X(i)-A_{N})/B_{N}\) of dynamical variables \(X(i)\) generated
by chaotic systems \(X_{j+1}(i)=
f^{(2)}_{\alpha=1.5,\delta_{1}=1,\delta_{2}=0.5}[X_{j}(i)]\),
 with \(N\)
different initial conditions \(X_{0}(i)|_{i=1,\cdots,N}\)
are plotted for
\(N=1,10,100,1000\) and \(10000\). In this case, the limit density converges to the asymmetric
L\'evy's stable law with the indices \(\alpha=1.5\)  and {\bf \(\beta=-\frac{9-4\sqrt{2}}{7}\)}.\\[1cm]

\begin{thebibliography}{99}
\bibitem{beta} When \(\beta=\pm 1\), these cases correspond to the 
purely asymmetric stable distributions over the semi-infinite support 
\(I_{+}=(0,\infty)\) for \(\beta=1\) and  \(I_{-}=(-\infty,0)\) 
for \(\beta=-1\). These completely 
asymmetric stable 
distributions can be generated by the chaotic systems 
\(X_{n+1}=|f_{\alpha,\delta}(X_{n})|\) with the exact densities 
\(\rho^{+}_{\alpha}(x)=\frac{2\alpha}{\pi}\frac{x^{\alpha-1}}{1+x^{2\alpha}}\)
on \(I_{+}\) for \(\beta=1\) and  
\( X_{n+1}=-|f_{\alpha,\delta}(X_{n})|\) with the density 
\(\rho^{-}_{\alpha}(x)=\frac{2\alpha}{\pi}\frac{|x|^{\alpha-1}}
{1+|x|^{2\alpha}}\) on \(I_{-}\) for \(\beta=-1\).
\end{thebibliography}
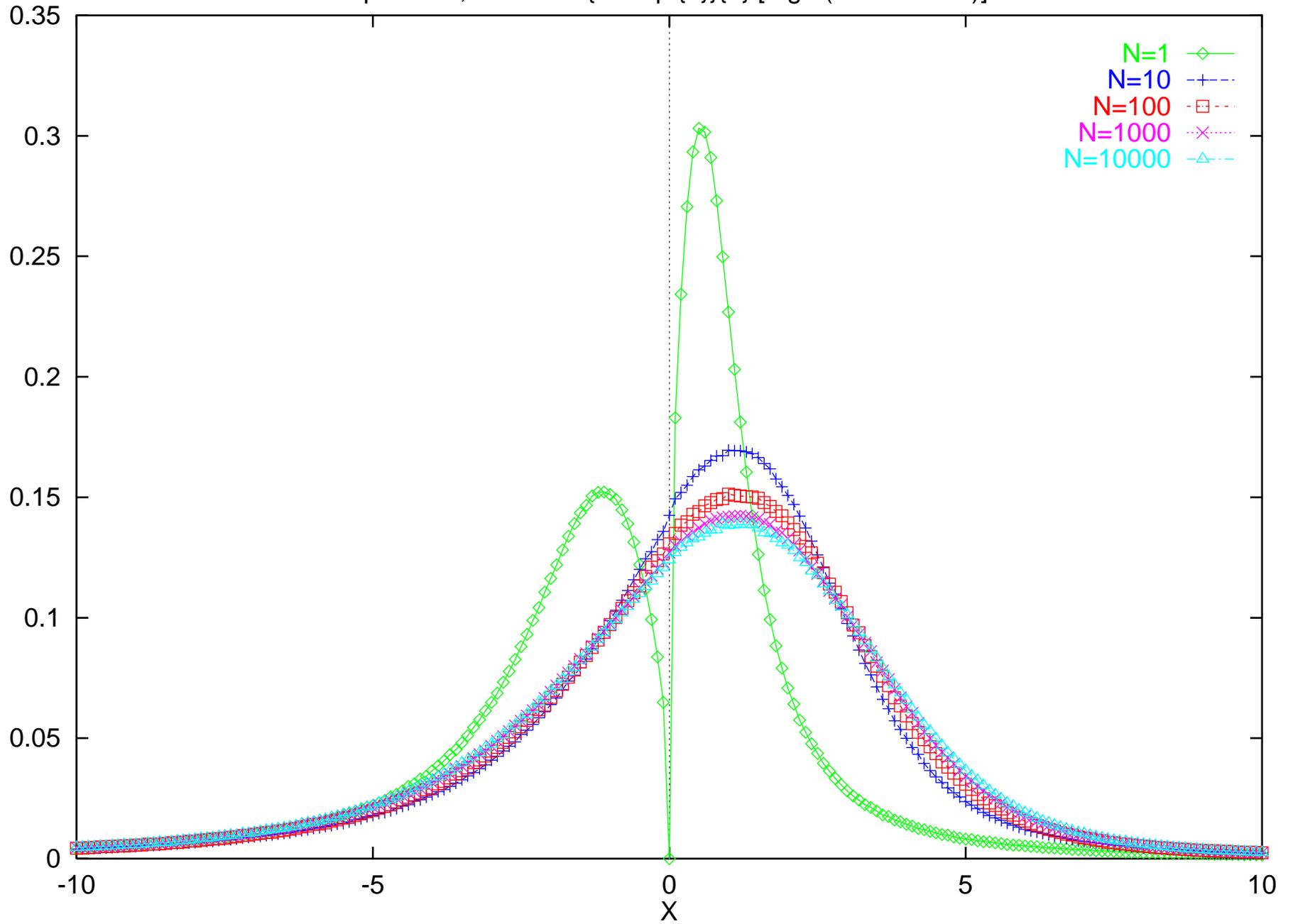
\end{document}